\documentclass[prb,%
citeautoscript,%
amsmath,amssymb,floatfix,reprint, superscriptaddress]{revtex4-1}
 
\usepackage[utf8]{inputenc}
\usepackage{hyperref}
\usepackage{amsthm}
\usepackage{amsmath}
\usepackage{natbib}
\usepackage{subfigure}
\usepackage[table]{xcolor}
\usepackage{graphicx}
\usepackage{dcolumn}
\usepackage{bm}
\usepackage{adjustbox}

\usepackage{enumitem}

\usepackage{siunitx}
\usepackage{mathptmx, microtype}

\begin{document}

\title{Pressure dependence of atomic dynamics in barocaloric ammonium sulfate: I. Rotations}

\author{Bernet E. Meijer}
\affiliation{School of Physics and Astronomy, Queen Mary University of London, Mile End Road, London E1 4NS, UK}
\author{Guanqun Cai}
\affiliation{School of Physics and Astronomy, Queen Mary University of London, Mile End Road, London E1 4NS, UK}
\author{Franz Demmel}
\affiliation{ISIS Neutron and Muon Source, Rutherford Appleton Laboratory, Chilton, Didcot, OX11 0QX, UK}
\author{Helen C. Walker}
\affiliation{ISIS Neutron and Muon Source, Rutherford Appleton Laboratory, Chilton, Didcot, OX11 0QX, UK}
\author{Anthony E. Phillips}
\affiliation{School of Physics and Astronomy, Queen Mary University of London, Mile End Road, London E1 4NS, UK}

\begin{abstract}\noindent 
Solid-state cooling using barocaloric materials is a promising avenue for eco-friendly, inexpensive and highly efficient cooling.
To design barocaloric compounds ready for deployment, it is essential to understand their thermodynamic behaviour under working conditions. 
To this end, we have studied the rotational dynamics in the molecular-ionic crystal ammonium sulfate under pressure, providing detailed insight into the origin of its large barocaloric effect. 
Using quasielastic neutron scattering experiments, we show that rotation of the ammonium cations is facilitated by pressure in the low-entropy phase, with the rotational ``hopping'' motion increasing in frequency as the pressure-induced phase transition is approached. 
%
We explain this unusual behaviour in terms of the competing hydrogen-bond networks represented by the two phases.
This work includes the first results of a recently developed low-background, high-pressure gas cell for neutron scattering, showcasing its power in obtaining high-precision measurements of molecular dynamics under pressure.
\end{abstract}

\maketitle
 
\date{\today}

\section{Introduction}

From computing to air conditioning to vaccine transport, cooling is crucial to modern society. 
However, with current techniques contributing to both ozone depletion and global warming, we are in desperate need for environmentally-friendly alternatives. 
%
A promising route is solid-state cooling, which avoids the use of ozone-depleting or greenhouse gases.
This technique uses the caloric effect -- an entropy change in response to an external field -- to create a heat pump.\citep{Moya2014}
%
%
A particularly appealing group of candidate materials are barocalorics, which can be switched between high- and low-entropy states by applying pressure.
From a practical standpoint, pressure is easy to apply in an industrial setting, without the need for large magnets (in the case of magnetocalorics) or the risk of material breakdown (in the case of electrocalorics)\citep{Usuda2017}.
%
Moreover, many barocaloric materials are cheap, non-toxic and constructed from simple and abundant building blocks.
Because the high-entropy state can have many different physical origins, including orientational, vibrational, electronic, and configurational disorder, barocalorics are found among many different classes of materials.\citep{Boldrin2021}
Examples include shape-memory alloys\citep{Manosa2010}; molecular and molecular-ionic crystals\citep{Li2019, Lloveras2019, Aznar2020, Li2020, Aznar2021}; spin-crossover complexes;\citep{Vallone2019, Romanini2021} and both inorganic\citep{Kosugi2021} and molecular perovskites\citep{Bermudez-Garcia2017, Salgado-Beceiro2020}. Such variety offers substantial hope that many more barocaloric materials are yet to be discovered.
%
However despite this great potential, the atomic understanding of the barocaloric effect that will be necessary to tune and even design barocaloric properties is still in its infancy.

In this work, we study the barocaloric effect of the inorganic salt ammonium sulfate (NH\textsubscript{4})\textsubscript{2}SO\textsubscript{4} through quasielastic neutron scattering (QENS) under pressure. 
Ammonium sulfate undergoes a solid-solid phase transition from the low-entropy phase with space group $Pna2_1$ to the high-entropy phase with space group $Pnam$\footnote{As has become conventional, we use a non-standard setting of space group no. 62, $Pnma$, so that the lattice vectors have the same direction in both phases.} at temperature $T_c =\SI{224}{K}$ (on heating) \citep{Schlemper1966, Lloveras2015}.
Because of a relatively rigid, low-density network of hydrogen bonding in the low-entropy phase, the unit cell volume \emph{decreases} on heating through the phase transition. As a result,
$T_c$ decreases with pressure, making this material an inverse barocaloric\citep{Lloveras2015}. 
%
The dynamics in ammonium sulfate have previously been studied by spectroscopic techniques including QENS\citep{Dahlborg1970, Goyal1978, Goyal1990} and NMR.\citep{OReilly1967, Kydon1969} 
Since QENS is sensitive to both the geometry and timescale of molecular reorientation, it is a particularly suitable means to study both structural and dynamic contributions to the entropy change. 
The experimental study of molecular dynamics under pressure, however, has only recently been made possible by the development of low-background gas cells\citep{Kibble2020}. We have taken this opportunity to study the rotational dynamics using quasielastic neutron scattering under pressure (this work) and the phonons using inelastic neutron scattering under pressure\citep{Yuan}.

This paper is organised as follows. We outline the experimental method in section \ref{sec:method} and give a theoretical background to quasielastic neutron scattering in section \ref{sec:theory}. In section \ref{sec:motion} we discuss the possible dynamical models of rotation in ammonium sulfate. The most reliable way to determine the appropriate dynamical model is an experiment without pressure cell, in which the signal-to-noise ratio is higher. The results of this experiment are presented in section \ref{sec:resultsA}. In section \ref{sec:resultsB} the confirmed dynamical model is then used to interpret the remarkable pressure dependence of the ammonium rotations. In section \ref{sec:resultsC} we conclude the study by emphasising the role of hydrogen bonds in this barocaloric's unique behaviour.


\section{Methods} \label{sec:method}

Neutron scattering experiments were performed on the OSIRIS spectrometer at the ISIS Neutron and Muon Source, U.K. OSIRIS is an indirect-geometry, time-of-flight spectrometer ideal for performing high-resolution quasielastic scattering experiments. 
Two different sample cells were used inside a closed cycle refrigerator (ccr): 1) a thin-walled aluminium can for ambient pressure measurements \citep{Meijer2021} and 2) a pressure cell \citep{Phillips2019}. 

OSIRIS uses a pyrolytic graphite analyser: the 002 reflection (pg002) gives an energy resolution function with FWHM 25.4 $\mu$eV and dynamic range of $\pm 0.5$ meV; the 004 reflection (pg004) gives energy resolution with FWHM 99.0 $\mu$eV and dynamic range of $-3.0$ to $4.0$ meV. The ambient pressure experiment made use of both analysers, giving a total $Q$-range of 0.18 to 3.6 \AA\textsuperscript{-1}. In the second experiment only the 002 reflection was used, resulting in a smaller $Q$-range of 0.18 to 1.8 \AA\textsuperscript{-1}.
%

For the first experiment, approximately 3 g of ammonium sulfate (purchased from Sigma Aldrich) was ground to a powder, enclosed in an aluminium foil sachet and then loaded into a cylindrical aluminium can (diameter: 24 mm, height: 45 mm) to give an approximately annular scattering geometry. This resulted in a mean sample thickness of approximately 0.9 mm and a similar scattering power as used in previous work \citep{Goyal1990}.

The pressure experiment employed a newly commissioned gas pressure cell\citep{Kibble2020}, manufactured from Ti-6Al-4V alloy (TAV-6). This bi-phase alloy is stronger than previously used TiZr, allowing thinner walls to contain a given pressure. For this reason and because the TAV-6 alloy inherently gives low noise, this cell has an exceptionally low background. The alloy does produce parasitic Bragg peaks starting at around $Q=2.5$ \AA{}\textsuperscript{-1}; they are however not visible in the $Q$-range of this experiment.
The cell has a cylindrical geometry, with walls of 5.3 mm and inner bore with a 7.0 mm diameter. The large sample thickness can cause significant multiple scattering in strongly scattering samples. To minimise this, the ammonium sulfate sample was diluted with KBr at a 1:16.2 mass ratio, resulting in approximately 18\% scattering. KBr has a very low incoherent scattering length compared to the ammonium ion, and the coherent cross section of KBr does not affect the neutrons from the low-energy 002 analyser setting; so this dilution does not interfere with the quasielastic experiment.

Empty can measurements (for background subtraction) and a low-temperature measurement (to obtain the experimental resolution function) were perfomed using the same experimental setup. Data collection was performed at temperatures of 200 K, 225 K, 300 K and 320 K, at pressures between 0 and 4.8 kbar. Elastic window scans of the elastic intensity as a function of temperature (supplementary figures S1 and S2) at 0 and 4 kbar confirmed that dynamics in this sample are visible on OSIRIS above about 200 K. 

Data reduction was performed in Mantid\citep{Arnold2014}. The pressure cell causes a background signal over all wave vectors due to the incoherent scattering of vanadium. Therefore, a proper cell subtraction is necessary to extract reliable intensities of the elastic contribution for the analysis of the elastic incoherent structure factor; this subtraction, including absorption corrections, was performed using the Paalman-Pings formalism\citep{Paalman1962} implemented in Mantid. All further data analysis was done in python.
\citep{Meijer_simpleQENS_2021}.


\section{Theoretical background} \label{sec:theory}

\begin{figure}
    \centering
    \includegraphics[width=1.0\linewidth]{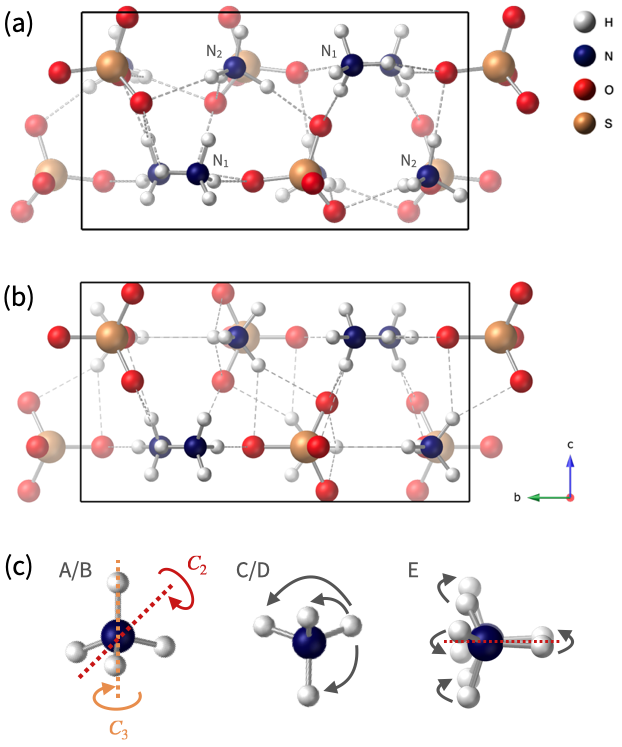}
    \caption{Crystallographic structure of ammonium sulfate in the \textit{Pna}2\textsubscript{1} (a) and \textit{Pnam} (b) phase\citep{Schlemper1966}. In both phases, there are two crystallographically distinct ammonium cations, indicated in (a) by N\textsubscript{1} and N\textsubscript{2}. (c) Dynamical models of ammonium reorientations as described in the main text.}
    \label{fig:fig1}
\end{figure}

Quasielastic scattering is low-energy inelastic scattering and appears as a broadening of the elastic line at $E=0$. Measuring this scattering is a powerful way to study stochastic rotational or translational diffusion \citep{Bee1988}.
%
The quantity of interest in a QENS experiment is the incoherent scattering function $S(Q, \omega)$ as a function of momentum transfer $Q$ and energy transfer $\omega$. The incoherent scattering length of $^1$H is much larger than the (in)coherent scattering length of any other atom; therefore, the incoherent single-particle hydrogen dynamics dominate the signal.
%
%
The theoretical incoherent scattering function can be related to the reduced experimental spectra $S_{\text{exp}}(Q, \omega)$ via
\begin{align}
    S_{\text{exp}}(Q, \omega) = R(\omega) \otimes \left( C(Q) S(Q, \omega) \right) + B(Q, \omega).
\end{align}
The factor $C(Q)$ includes the Debye-Waller factor  $e^{-\langle u^2 \rangle Q^2}$, where the mean-square displacement $\langle u^2 \rangle$ represents molecular vibrations and librations. $C(Q)$ was allowed to vary freely with $Q$ in the global fit.\footnote{The mean-square displacement was determined from the Debye-Waller factor using an elastic window scan; however, the results were noisy and not of interest for this study.}
The model is convolved with the experimental resolution function $R(\omega)$ (obtained from a measurement at 10 K) to account for the energy resolution of the spectrometer.
Finally, a background term $B(Q, \omega)$ is added to account for inelastic 
contributions to the measured signal, which can be assumed to be linear in this small energy transfer range.

On the OSIRIS timescale, we expect to probe rotational rather than translational diffusion of hydrogen atoms. Translational diffusion of \textit{whole} ammonium cations would be a high-energy barrier process not visible on the OSIRIS timescale; translational diffusion of \textit{individual} hydrogen atoms between cations is highly unlikely since the cations lie far apart.
The conclusion that ammonium cations undergo solely jump-reorientations is furthermore consistent with both previous NMR \citep{OReilly1967, Kydon1969} and QENS studies.\citep{Dahlborg1970, Goyal1978, Goyal1990}
The scattering function for rotational dynamics can be written as an elastic component represented by a delta function, and a sum of Lorentzian-shaped quasielastic contributions\citep{Bee1988}:
\begin{align}
    S(Q, \omega) = A_0(Q) \delta(\omega) + \sum_{i=1}^{n} A_i(Q) \frac{1}{\pi} \frac{\frac{1}{2}\Gamma_i}{\omega^2 + (\frac{1}{2}\Gamma_i)^2},
\end{align}
where $\sum_{j=0}^{n} A_j = 1$. Each dynamical process probed is represented by a Lorentzian function, whose full-width at half-maximum $\Gamma_i$ is inversely proportional to the characteristic time $\tau_i$ of that process: $\tau_i=2\hbar/\Gamma_i$. All of the models we consider here are jump-rotational, whereby  hydrogen atoms hop instantaneously between equilibrium sites. In this context the characteristic time $\tau_i$ is the mean residence time between jumps; its inverse is the rotational jump frequency.
In this case of localised rotational motion, $\tau_i$ is independent of momentum transfer $Q$.

The amplitude $A_0(Q)$ of the elastic signal is called the Elastic Incoherent Structure Factor (EISF). It is defined as the proportion of elastic scattering to the total scattered intensity:
\begin{align}
    A_0(Q) = \frac{I\textsubscript{elastic}(Q)}{I\textsubscript{elastic}(Q)+I\textsubscript{inelastic}(Q)}.
\end{align}
Its $Q$-dependence is determined by the geometry of the localised motion of the individual hydrogens. We will present hypotheses for the geometric model and their corresponding theoretical EISFs in section \ref{sec:motion}. 
In order to relate the model EISFs to the experimentally measured intensity $A_0(Q)$, we need to take into account the fact that a QENS signal consists of two contributions. One part of the signal originates from molecules that are mobile on the timescale of the spectrometer, while another part originates from molecules that appear to be static on that timescale\citep{Songvilay2019}:
\begin{align}
    I_{\text{tot}} = fI_{\text{free}} + (1-f)I_{\text{bound}},
\end{align}
where $f$ is the fraction of ammonium cations that are free to rotate. This means that the total experimental EISF can be written as
\begin{align}
    A_0(Q) = (1-f) + f \times \text{EISF}_{\text{geometric}}.
\end{align}


\section{Dynamical models} \label{sec:motion}

\begin{table}  
\begin{tabular}{ m{1.0cm} m{6.5cm}  } 
\toprule
Model& EISF\textsubscript{geometric} \\
\colrule
A/B&  $A_0(Q) = \frac{1}{4}\left[4-2\left(1-j_0(Qd_{\text{H-H}})\right)\right]$ \\
C & $A_0(Q) = \frac{1}{4}\left[4-3\left(1-j_0(Qd_{\text{H-H}})\right)\right]$ \\
D& $A_0(Q) = \frac{1}{4}\left[4-2.75\left(1-j_0(Qd_{\text{H-H}})\right)\right]$\\
E & $A_0(Q) = \frac{1}{8} \sum_{i=1}^{8} \frac{1}{4}\left[4-2\left(1-j_0(Q\tilde{d}_i)\right)\right]$ \\
\botrule
\end{tabular}
\caption{EISF for each model. Models A-C are derived using the group-theoretical formalism of Thibaudier and Volino \citep{Thibaudier1973,Thibaudier1975,Bee1988}. Model D was proposed by Goyal and Dasannacharya\citep{Goyal1978}. In model E, the signal is the average of the two-site reorientational signal \citep{Bee1988} of each of the eight distinct hydrogen atoms. $d_{\text{H-H}}$=1.64 \AA{} is the distance between hydrogen atoms in the ammonium tetrahedron\citep{Schlemper1966}; $\tilde{d}_i \in \{0.15, 0.10, 0.50, 0.50, 0.57, 0.52, 0.82, 0.82\}$ is the distance of hydrogen atom $i$ between the two disordered sites.  \label{tab:eisf}}
\end{table}

Ammonium sulfate has two crystallographically distinct ammonium cations in the unit cell. In the high-symmetry phase, it has been shown that the reorientational timescales of the two types are similar \citep{OReilly1967, Goyal1978}. This is also the case in the low-symmetry phase down to at least 215 K \citep{Goyal1990}. It is only at very low temperatures ($T < 170$ K) that NMR experiments could separate out two different timescales \citep{OReilly1967}. For the phase points in the current experiment, we therefore expect to approximate the signal with a single effective timescale, with the possible exception of the lowest-temperature data set at \SI{200}{K}. 

Regarding the geometry of the motion, we will consider five different dynamical models:
\begin{enumerate}[label=\Alph*]
    \item 180\textsuperscript{$\circ$} jumps about \textit{one} of three 2-fold rotation axes bisecting the H-N-H angles;
    \item 120\textsuperscript{$\circ$} jumps about \textit{one} of four 3-fold rotation axes along the N-H bonds;
    \item A four-site tetrahedral jump model, where each H atom visits all four sites of the ammonium cation with equal probability. This model can be achieved either by jumps about \textit{all} of the 3-fold rotation axes, by jumps about \textit{all} of the 2-fold rotation axes or by jumps about all 3-fold and 2-fold rotation axes;
    \item One of the ammonium cations reorients according to model C; the H atoms of the second ammonium cation visit all 4 sites, but not with equal probability. This model was proposed by Goyal and Dasannacharya\citep{Goyal1978}.
    \item Ammonium cations hop between two disordered states that are each other's mirror image in the mirror plane at $z=1/4$. One state is the ammonium position in the \textit{Pna}2\textsubscript{1} phase; the second is produced by reflecting those positions in the mirror plane. Such a reflection cannot occur physically without the ammonium ion distorting during the ``hop''.
\end{enumerate}

These motions are illustrated in figure \ref{fig:fig1}(b) and mathematical expressions for the EISF are given in table \ref{tab:eisf}. Note that models A and B have identical EISFs because of tetrahedral symmetry. This experiment therefore cannot distinguish between those two dynamical models. 


While models A--C are possible for all molecules with tetrahedral symmetry, model D was formulated specifically for ammonium sulfate.~ Goyal and Dasannacharya\citep{Goyal1978} found that the motion was best described by a mathematical model that is a mix between models A, B and C. The physical interpretation of this is that rotations around the different $C_2$ and $C_3$ axes are all possible, but not equally probable. They propose that the EISF is given by the functional form
\begin{align} \label{eq:EISF_Goyal}
    \text{EISF}_{\text{geometric}}(Q) = \frac{1}{4}\left[4-n\left(1-j_0(Qd_{\text{H-H}})\right)\right],
\end{align}
where $j_0(x)=\sin(x)/x$ is the spherical Bessel function of zeroth order. If we take $n = 2$, this reduces to the formula expected for models A and B, while if $n = 3$ it reduces to the formula expected for model C. However, it is also possible to let $n$ vary as a parameter, and Goyal and Dasannacharya\citep{Goyal1978} found that a reasonable fit to the data was obtained with $n = 2.75$; we take this specific value as our model D.


Of these five models, A--D are all consistent with the ammonium ion's tetrahedral symmetry in an ordered crystal structure, while model E requires crystallographic disorder about the mirror plane. In fact, as we have discussed elsewhere, there is no evidence for such disorder.\citep{Meijer}
Nonetheless, because order-disorder models have been discussed at length in the literature,%
\citep{OReilly1967, OReilly1969, De1994} for completeness we include this model in our hypotheses in order to confirm that we can indeed rule out such a mechanism based on spectroscopic as well as crystallographic evidence.



\begin{figure}
    \centering
    \includegraphics[width=1.0\linewidth]{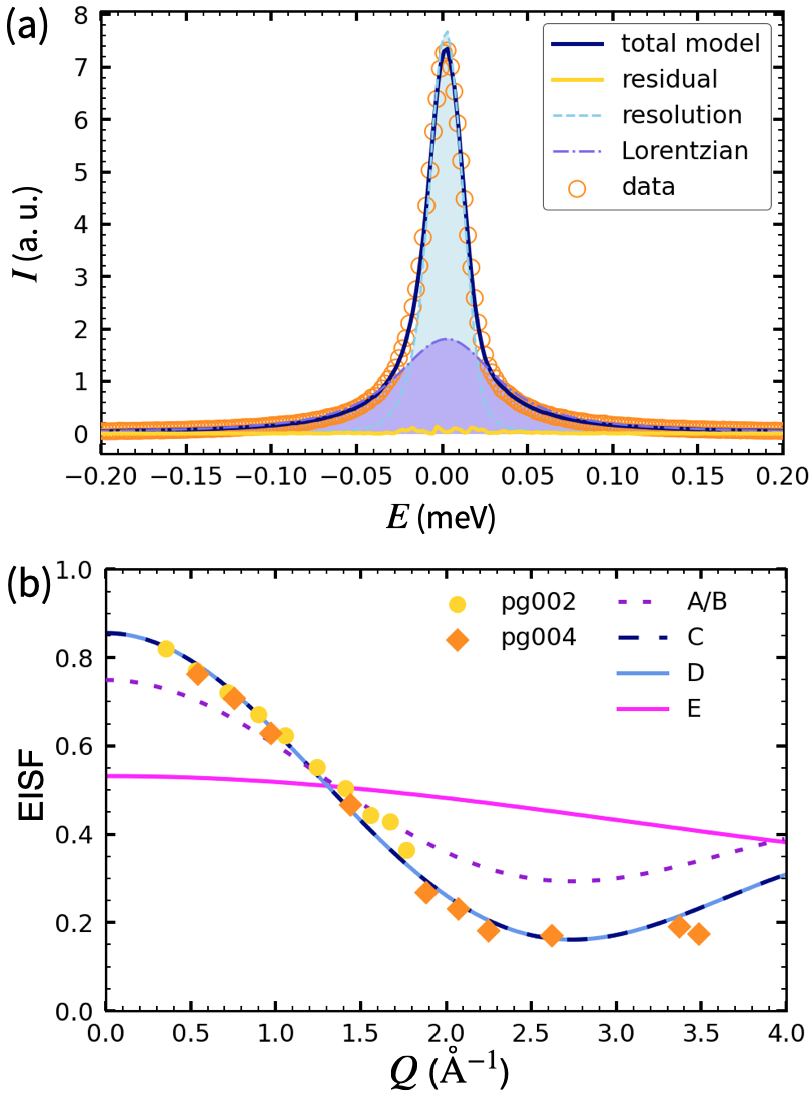}
    \caption{QENS results for ammonium sulfate at ambient pressure. A representative fit at 200K, $Q=0.56$ \AA\textsuperscript{-1} (a) shows that the data is fitted well with a single Lorentzian. Fitted models to experimental EISF at 300K show that models C and D, described in the text, are in good agreement with the data.}
    \label{fig:nonP_qens}
\end{figure}

\section{Results and Discussion} \label{sec:results}

\subsection{Determining dynamical model} \label{sec:resultsA}

To interpret the pressure dependence of ammonium sulfate's dynamics, it is necessary to first determine the appropriate rotational dynamical model (section \ref{sec:motion}) using the data at ambient pressure, where background noise is low.
%
At all temperatures, a model with a single Lorentzian was sufficient to fit the data within experimental error.
%
A representative fit at 200 K is shown in figure \ref{fig:nonP_qens}(a). This suggests that the reorientations of both crystallographically distinct cations occur on a single timescale within the precision of this experiment, even down to 200 K. 

At all temperatures, the fitted width is independent of momentum transfer $Q$ (see figure S3), indicating that the molecules undergo localised rotational motion rather than diffusive motion. This agrees with the previous QENS studies, and confirms the consideration in section \ref{sec:theory} that there is no physically plausible mechanism for diffusion in this material.
In all subsequent fits, we therefore fixed the Lorentzian width $\Gamma$ as a global parameter independent of $Q$.

\begin{table}
    \centering
    \begin{tabular}{SSS|SS}
    \toprule
    {$T$ (K)} & {PG002 (ps)} & {PG004 (ps)} & {$T$ (K)} & {Goyal \emph{et al.} (ps)} \\
    \colrule
    200 & 18.2(4) & 14.31+-0.07 & 215 & 20.8(32) \\
    225 & 7.43(3) & 5.635+-0.012 & 230 & 13.1(40) \\
    300 & 3.15(4) & 2.840+-0.010 & 300 & 2.13(20) \\
    \botrule
    \end{tabular}
    \caption{Mean residence times $\tau$ as determined from our own data are in reasonable agreement with literature results, given the different experimental resolutions.\cite{Goyal1978, Goyal1990} Note that Goyal et al.'s definition of the residence time differs from ours by a factor of $1/2\pi$; for ease of comparison, the residence times quoted here are transformed back to our own definition.}
    \label{tab:tau}
\end{table}

%
The measured mean residence times are given in Table~\ref{tab:tau}. 
%
The differences between the measured timescales in our PG002 setting, our PG004 setting and measurements by Goyal et al.\citep{Goyal1978, Goyal1990} are due to the different experimental resolutions of 25$\mu$eV, 99$\mu$eV and 15$\mu$eV respectively. Nevertheless, the timescales are roughly in the same range, and confirm in particular that the motion speeds up on heating.

We now consider the geometry of the motion. The experimental EISF at 300 K is shown in figure \ref{fig:nonP_qens}(b). (Spectrum fits and EISFs at the other temperatures are shown in supplementary figure S4.) We note that at all temperatures, the EISF does not quite reach 1 as $Q\rightarrow 0$, which is likely to arise from multiple scattering. Similar behaviour was seen in previously reported work\citep{Goyal1978, Goyal1990}, where a similar sample thickness was used. In order to fit the data to the geometric models, it was therefore necessary to account for multiple scattering by adding a further parameter $m$\citep{Songvilay2019}:
\begin{align} \label{eq:eisf}
    A_0(Q) = (1-f) + fm \times \text{EISF}_{\text{geometric}}.
\end{align}
In the case of no multiple scattering, $m$ should refine to 1; if multiple scattering is significant $m$ refines to values less than 1. 

For each temperature, the experimental $A_0(Q)$ was fitted to each of the geometric EISFs in table \ref{tab:eisf} using equation \ref{eq:eisf}, refining the parameters $f$ and $m$ using a least-squares method (results are listed in supplementary tables S1-3). The representative fit at 300 K of figure \ref{fig:nonP_qens}(b) shows that models C and D are in equally good agreement with the data. Note that, as discussed in section \ref{sec:motion}, the empirical parameter $n$ of eqn. \ref{eq:EISF_Goyal} is \textit{fixed} at $n=3$ and $n=2.75$ for models C and D respectively. In this fit these EISFs nevertheless have the same shape due to different fitted values of $f$ and $m$ (see supplementary tables S1-3).

Compared to model C, model D has an extra emprically determined parameter $n$. By applying Ockham's razor it follows that the ``tetrahedral tumbling'' model C best describes the data at both 225 K and 300 K. 
Although the 200 K data could not distinguish between models A/B, C and D, we posit that the tetrahedral tumbling model is the most plausible at this temperature too. We will support this claim using the pressure data in the next section. Reorientational dynamics according to model C have been observed in other ammonium salts including ammonium perchlorate\citep{Prask1975}, ammonium borohydride\citep{Andersson2020} and ammonium cyanate\citep{Desmedt2008}.



\subsection{Pressure dependence} \label{sec:resultsB}

\begin{figure*}
    \centering
    \includegraphics[width=\textwidth]{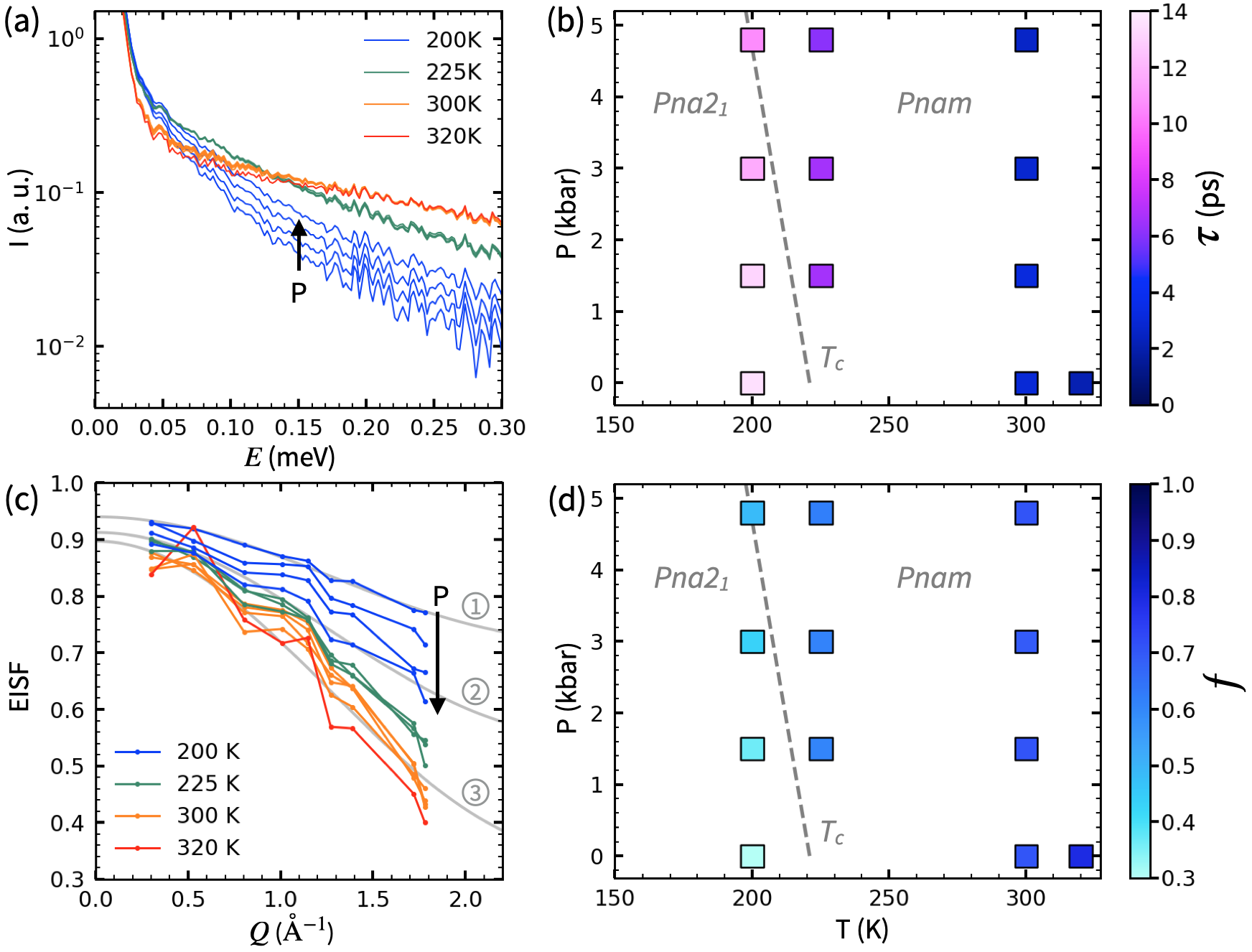}
    \caption{QENS results for ammonium sulfate under pressure. (a) Detail of fits to the experimental spectra, summed over $Q$, at different phase points. The fits are noisy due to the convolution with the experimentally determined energy resolution. At 200 K, the tails of the signal broaden with pressure, indicating a decrease in mean residence time $\tau$ with pressure: this can be seen clearly in a phase diagram where the phase points are coloured by the fitted value of $\tau$ (b). (c) Experimental EISF at different phase points. The grey lines are fits to EISF model C, at the phase points (200 K, ambient pressure) (1), (200 K, 4.8 kbar) (2) and (300 K, 4.8 kbar) (3). EISF fits at all phase points are plotted in supplementary figures S5-8. The corresponding fitted values of $f$, the fraction of ammonium cations free to rotate, at all phase points are presented in a phase diagram (d).}
    \label{fig:fig3}
\end{figure*}

As in the ambient pressure experiment, the data from the pressure cell was best fitted to a single-Lorentzian model. %
Again, the Lorentzian width was constrained to be $Q$-independent. All spectrum fits and fits to the EISF are shown in supplementary figures S5-8.

As seen in figure \ref{fig:fig3}(c), the shape of the resulting $A_0(Q)$ is the same at all temperatures and pressures, apart from an overall scale factor which can be explained by varying the fraction of rotating cations $f$. This shows that the geometry of the motion is the same at all temperatures and pressures.
This is consistent with the results of Goyal \emph{et al.}\cite{Goyal1990}, who showed that there is no qualitative change in the geometry of the motion above and just below the phase transition. 
Hence, since the ambient pressure experiment showed that the ammonium cations reorient according to model C at 225 and 300 K, this model is likely also to hold down to 200 K.

The experimental $A_0(Q)$ was consequently fitted to EISF model C, refining the parameters $m$ and $f$ at each phase point. The results are summarised in figure \ref{fig:fig3}: for each measured phase point, the fitted mean residence time $\tau$ (\ref{fig:fig3}(b)) and the fraction of rotating ammonium cations $f$ (\ref{fig:fig3}(d)) are shown. In the \textit{Pna}2\textsubscript{1} phase both the jump rate and the fraction of rotationally activated cations increase with pressure as the material approaches the phase transition. Once in the \textit{Pnam} phase, the jump rate and fraction of activated cations remain approximately constant under further pressurisation. 

We note that the phase point (200 K, 4.8 kbar) lies very close to the phase transition. The phase coexistence curve sketched in the graph is taken from isobaric measurements\citep{Lloveras2015} and only serves as an approximation to the phase transition temperature for our isothermal approach to this phase point. However, we can estimate which is the correct crystallographic phase at (200 K, 4.8 kbar) by looking at the EISF (figure \ref{fig:fig3}(c)). When going from this phase point to 225 K, the EISF is lowered substantially, corresponding to a large jump in $f$. Between 225 K and 300 K on the other hand, despite the big change in temperature, the EISF only changes slightly. Hence, the relatively big jump in $f$ between (200 K, 4.8 kbar) and 225 K might be explained by the phase transition, which would suggest that this point lies in the \textit{Pna}2\textsubscript{1} phase.

The pressure dependence in the \textit{Pna}2\textsubscript{1} phase is remarkable. It is common for phonon frequencies to increase with pressure, as a decrease in volume forces the atoms closer together and increases the interatomic force constants \citep{Dove1993}.
At first glance, one might similarly expect that a decrease in volume will raise the energy barriers between neighbouring hydrogen sites, therefore reducing the probability of jump-reorientations. One would consequently expect a decrease in the frequency of jump-reorientations with pressure. 
This is for example the case in the barocaloric molecular crystal neopentylglycol (NPG), as shown in a recent QENS study\citep{Li2019}.
However, in ammonium sulfate's \textit{Pna}2\textsubscript{1} phase, increasing pressure has the opposite effect on the energy landscape. Figures \ref{fig:fig3}(b, d) show that molecular reorientations are in fact facilitated by increasing pressure. 
This finding is supported by inelastic neutron scattering measurements to be reported in our next paper, showing that most of the librational phonon modes of the ammonium cations soften with pressure, indicating a negative Gr\"uneisen parameter\citep{Yuan}. 
This behaviour testifies to a flattening of the energy landscape with pressure. In a flatter energy landscape, it is more likely for both ammonium librations and ammonium jump-rotations to occur: the librational phonon frequency \emph{decreases} while the jump-rotational frequency \emph{increases}.


\subsection{Competing hydrogen bond networks} \label{sec:resultsC}

An explanation for the flattening of the energy landscape can be found in the competing arrangements of hydrogen bonds. In the \textit{Pna}2\textsubscript{1} phase, the ammonium cations are oriented in such a way to create strong, linear N–H···O bonds with an average length of 2.13 \AA{} (see figure \ref{fig:fig1} and also the detailed hydrogen bond analysis by Malec et al.\citep{Malec2018a}). This results in a stiff hydrogen bond network and accounts for the relatively large unit cell volume of this phase. Increasing pressure draws the ions closer together and decreases the hydrogen bond distances beyond their equilibrium. 
Consequently, the energy minima of this arrangement are raised, and the hydrogen atoms are more likely to both slightly tilt sideways (an effect we call `buckling') and jump around to the next symmetry position (`hop'). 


In the \textit{Pnam} phase, the ammonium cations orient to form a symmetrical hydrogen bond arrangement with longer, weaker hydrogen bonds of 2.23 \AA{} average length (figure \ref{fig:fig1}) although the structure overall is more closely packed. 
The relatively flat energy landscape created by these weak hydrogen bonds allows for a high hopping frequency.
However, in this denser and more symmetrical structure, buckling of the hydrogen bonds no longer provides a way to relieve strain; and hence applying pressure does not destabilise the structure in the same way. Therefore, increasing pressure only has a small effect on the hopping frequency, while it increases the librational phonon frequencies (positive Gr\"uneisen parameter).

Ammonium sulfate's behaviour under pressure forms a stark contrast with that of other molecular barocalorics such as NPG, where the reorientational frequency drops with pressure, eventually causing the molecules to freeze into equilibrium positions.
This difference might be explained by these materials' difference in bonding.
In its low-temperature phase, NPG has one-dimensional chains of hydrogen bonds\citep{Nakano1969}; the remaining intermolecular interactions are van der Waals. In ammonium sulfate on the other hand, hydrogen bonds dominate the intermolecular interactions in all three dimensions. 
This suggests that, as a general heuristic, molecular reorientation frequencies in systems governed by van der Waals forces can be expected to decrease under pressure: as the atoms are forced closer together, the force constants increase and the barrier to rotation is raised. 
However when hydrogen bonding is the dominant interaction, even after applying pressure the van der Waals forces have a negligible effect on the barrier to rotation.
Instead, depending on the strength of the hydrogen bonds, increased pressure can destabilise the molecular arrangement and increase the hopping frequency.



\section{Conclusion} \label{sec:conclusion}

From high-resolution QENS data, we have determined the pressure and temperature dependence of ammonium rotations in barocaloric ammonium sulfate.
This study confirms that the geometry of the motion does not change over the phase transition, and can be described by a tetrahedral tumbling model in which the hydrogen atoms visit all four sites of the ammonium tetrahedron.
Importantly, there is no evidence for a 2-site jump model over a mirror plane. This is consistent with the absence of crystallographic disorder, and demonstrates that the large entropy change in this material cannot be explained by configurational entropy. We have previously discussed the implications for the elusive source of ammonium sulfate's entropy change.\citep{Meijer}

Our measurements under pressure reveal that the barrier to rotation in the low-temperature phase decreases with pressure, while in the high-temperature phase pressurisation has little effect. With pressure usually suppressing the reorientational dynamics, 
ammonium sulfate's behaviour is remarkable, and related to its inverse barocaloric behaviour. In the low-temperature phase, the stiff, sparse hydrogen bond network can be destabilised by applying pressure, which increases the probability for hydrogen atoms to buckle sideways, thereby decreasing the barrier to reorientation as the phase transition is approached. 
Conversely, in the high-temperature phase, the denser hydrogen bond network cannot respond to pressure by buckling in the same way, 
so that the barrier to reorientation changes rather less with increasing pressure in this phase.

In summary, this material's excellent barocaloric properties all originate in some way from the competing arrangements of hydrogen bonds. 
First, due to the rigidity of the hydrogen bond network in the low-entropy \textit{Pna}2\textsubscript{1} phase, there is a large volume change between the phases and hence pressure strongly induces the transition to the high-entropy \textit{Pnam} phase. This results in a large barocaloric coefficient $|dT_c/dP|$. This also makes ammonium sulfate an inverse barocaloric, in which the entropy increases with pressure. 
Second, the ease of sideways buckling of the hydrogen bonds in the low-entropy phase is reflected not only in the rotational hopping dynamics reported here and in the phonon spectrum\citep{Yuan}, but also in the high entropy change, which is responsible for the large refrigerant capacity\citep{Meijer}.
Our results suggest that molecular ions rich in hydrogen bond donors and acceptors may be suitable components for designing new materials with similarly competing hydrogen bond networks; and that such networks, while challenging to design, might be a fruitful route for producing new barocaloric materials.

\begin{acknowledgements}
Experiments RB1920740 and RB2000267 at the ISIS Neutron and Muon Source were allocated beamtime funded by the Science and Technology Facilities Council, and the authors thank C. Goodway and M. Kibble for experimental support. BEM thanks ISIS and Queen Mary University of London for PhD studentship funding. GC thanks the China Scholarship Council. HCW and AEP thank EPSRC for financial support: EP/S035923/1 and EP/S03577X/1. 
\end{acknowledgements}

\end{document}